\documentclass[aps, prb, amssymb, superscriptaddress]{revtex4-2} 

\usepackage{graphicx}
\usepackage{siunitx}
\usepackage{hyperref}
\usepackage{amsmath}
\usepackage{xcolor}

\begin{document}

\title{Gate control, g-factors and spin orbit energy of p-type GaSb nanowire quantum dot devices}

\author{Sven Dorsch}
\email{sven.dorsch@ftf.lth.se}
\affiliation{Solid State Physics and NanoLund, Lund University, Box 118, SE-221 00 Lund, Sweden}
\author{In-Pyo Yeo}
\affiliation{Solid State Physics and NanoLund, Lund University, Box 118, SE-221 00 Lund, Sweden}
\author{Sebastian Lehmann}
\affiliation{Solid State Physics and NanoLund, Lund University, Box 118, SE-221 00 Lund, Sweden}
\author{Kimberly Dick}
\affiliation{Solid State Physics and NanoLund, Lund University, Box 118, SE-221 00 Lund, Sweden}
\affiliation{Center for Analysis and Synthesis, Lund University, Box 118, SE-221 00 Lund, Sweden}
\author{Claes Thelander}
\affiliation{Solid State Physics and NanoLund, Lund University, Box 118, SE-221 00 Lund, Sweden}
\author{Adam Burke}
\email{adam.burke@ftf.lth.se}
\affiliation{Solid State Physics and NanoLund, Lund University, Box 118, SE-221 00 Lund, Sweden}

\date{\today}%

\begin{abstract}
Proposals for quantum information applications are frequently based on the coherent manipulation of spins confined to quantum dots. For these applications, p-type III-V material systems promise a reduction of the hyperfine interaction while maintaining large $g$-factors and strong spin-orbit interaction. In this work, we study bottom-gated device architectures to realize single and serial multi-quantum dot systems in Schottky contacted p-type GaSb nanowires. We find that the effect of potentials applied to gate electrodes on the nanowire is highly localized to the immediate vicinity of the gate electrode only which prevents the formation of double quantum dots with commonly used device architectures. We further study the transport properties of a single quantum dot induced by bottom-gating, find large gate-voltage dependent variations of the $g^*$-factors up to $8.1\pm 0.2$ as well as spin-orbit energies between $110$-$230\,\si{\mu eV}$.

\end{abstract}

\maketitle

A key challenge for spintronic and quantum electronic applications is achieving long coherence- and spin-lifetimes \cite{ZuticRMP2004}. Here, p-type materials can be beneficial in comparison to more conventional n-type systems \cite{BrunnerScience2009}. In the valence band, the p-orbital symmetry efficiently reduces the hyperfine interaction of free holes through a suppression of the contact term \cite{KolodrubetzScience2009,BrunnerScience2009,PrechtelNatMat2016,Pribiag2013electrical}.

Consequently, interest in p-type quantum dot (QD) devices, where the coherent manipulation of spins confined to the QD enables qubit operation \cite{LossPRA1998}, is growing fast and systems based on InGaAs \cite{PrechtelNatMat2016}, Ge \cite{ScappucciNatRevMat2020}, Si \cite{ZwanenburgRMP2013} or ambipolar InSb \cite{NadjPRL2012,Pribiag2013electrical} are widely regarded as promising candidates. Another, hitherto less investigated candidate is GaSb, which due to its high hole mobility and the expected strong spin-orbit interaction in III-V semiconductors offers interesting material properties \cite{KaralicPRB2019, BennettSSE2005}. Hole transport in GaSb QD devices has so far only been studied in GaSb/InAsSb core-shell \cite{GanjipourPRB2015} and plain GaSb nanowires, where metallic contacts at low temperatures form Schottky barriers allowing the formation of a QD between closely spaced electrodes \cite{GanjipourAPL2011}.

Schottky barrier defined QDs are fabrication-limited by the smallest achievable spacing between the contacts and the QD dimensions and tunnel rates vary dependent on the gate voltage and the applied bias. Smaller QD structures, however, are beneficial as they allow easier access to quantum confinement effects in the electronic structure \cite{GanjipourAPL2011}. Consequently, an important step toward the realization of GaSb nanowire based spintronic devices is the development of device architectures allowing the formation and characterization of small QDs. For the more common nanowire material InAs the development of local bottom-gates \cite{FasthNanoLett2005} and later epitaxially defined InP-InAs-InP \cite{BjorkPRB2005,FuhrerNanoLett2007} as well as Wz-Zb-Wz polytype structures \cite{NilssonPRB2016,BarkerAPL2019} have enabled small, tunable and flexible high quality QD and serial double quantum dot (DQD) devices, which are now widely used for various transport studies. For GaSb nanowires, epitaxially defined QD structures are not yet experimentally available and the design of flexible, local gate architectures is challenging for Schottky contacted nanowires. In this work we investigate to what extent control can be achieved in p-type GaSb:Zn nanowires by studying the transport properties of bottom-gate defined single- and multi-QDs.

\begin{figure*}
    \includegraphics{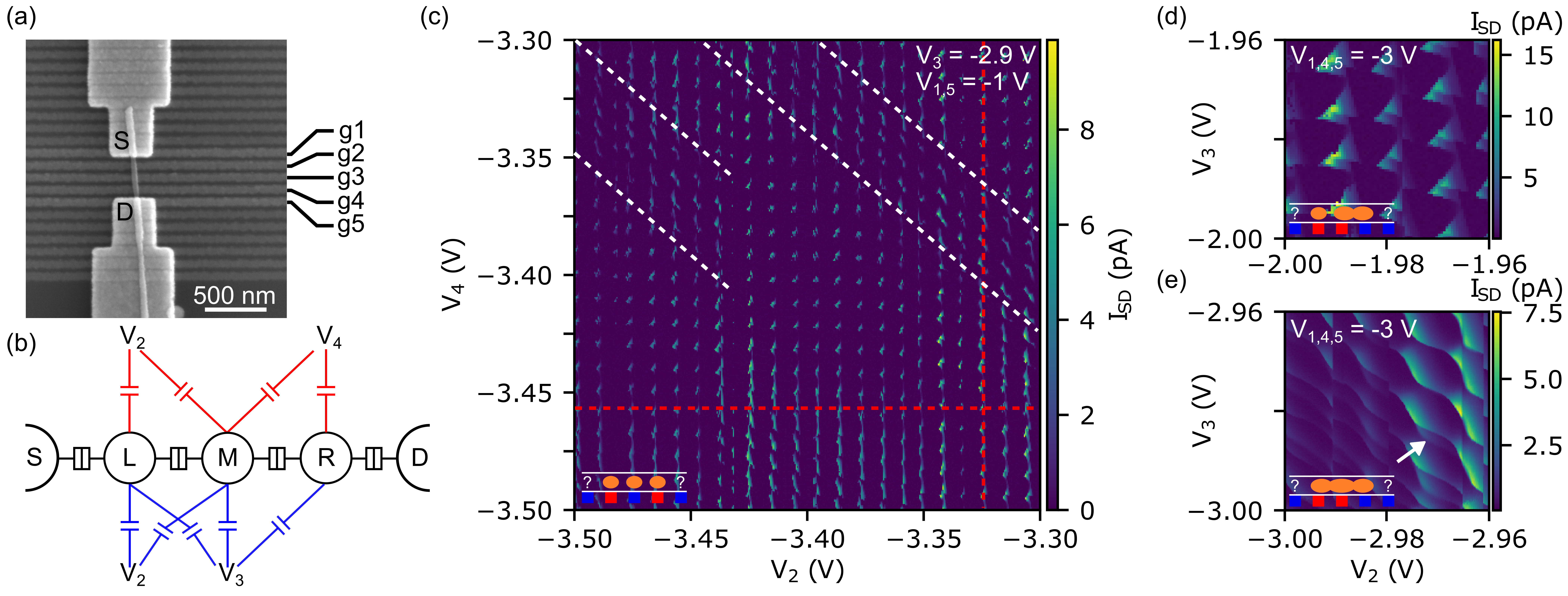}
    \caption{(a) Scanning electron microscope image of device A, a Schottky-contacted GaSb nanowire on five gate stripes g1 - g5. (b) Measurement configurations for panel c indicated in red and panels d/e in blue. (c) Charge stability diagram as a function of $V_2$ and $V_3$ at $V_{\mathrm{SD}}=1\,\si{mV}$. Pattern repetition axis are indicated. (d,e) Charge stability diagram at $V_{\mathrm{SD}}=3\,\si{mV}$ as a function of $V_2$ and $V_3$ in a weak- (d) and strong (e) coupling regime between the left and middle QD. $V_{\mathrm{BG}}=-10\,\si{V}$ for all measurements in Fig.~1.\label{fig1}}
\end{figure*}

Arrays of of gate electrodes were fabricated on top of a back-gated Si/SiO$_2$ substrate and covered by a layer of $\mathrm{HfO}_2$. Zinc-doped GaAs-GaSb:Zn nanowires were deposited on top of the gate arrays and the GaSb segment was contacted by Ni/Au Schottky contacts (source, drain). For device geometry A, shown in Fig.~\ref{fig1}(a), we use five underlying gates (g1-g5 with applied voltages $V_{1-5}$) with the purpose of forming a DQD. Here, g1 and g5 are located below the Schottky contacts with the aim to allow control over tunnel couplings to the contacts. Gates g2 and g4 are designed to act as plunger-gates to a left (L) and right (R) QD while g3 induces a barrier between the QDs.

Fig.~\ref{fig1}(c) shows a characteristic charge stability diagram of the device as a function of $V_2$ and $V_4$ while $V_3$ is held constant at $-2.9\,\si{V}$. We find finite bias triangles, arranged along horizontal and vertical symmetry axes (dashed red lines) with additional diagonal features (enclosed by white dashed lines). This behaviour is comparable to results obtained on a serial triple quantum dot (TQD) \cite{FroningAPL2018} and we explain this by the additional formation of a middle (M) QD atop g3, resulting in the measurement configuration indicated in red in Fig.~\ref{fig1}(b). To verify TQD formation, we next measure the charge stability diagram as a function of $V_2$ and $V_3$ with $V_{1,4,5}=-3\,\si{V}$. In this configuration, sketched in blue in Fig.~\ref{fig1}(b), we then expect to be able to control the interdot coupling between the left and middle QD by tuning the extent of the conductive islands within the nanowire. Indeed, for sufficiently low negative gate voltages $V_{2,3}$ we find behaviour matching that of a weakly coupled DQD in Fig.~\ref{fig1}(d). By decreasing the voltages $V_{2,3}$, shown in Fig.~\ref{fig1}(e), the charge stability diagram transitions resemble that of a strongly coupled DQD. We note that additional cells within Fig.~\ref{fig1}(e) (see for example white arrow) again indicate TQD formation.

We find transport across the device to only be possible if sufficiently negative voltages are applied to g2, g3 and g4 to form three QDs. For more positive voltages $V_3$, where no middle QD is formed, transport remains blocked which prevents the formation of a DQD. This, in combination with the formation of independently addressable QDs located on top of neighbouring gates, indicates that gate-action on the nanowire occurs only very localized by either inducing or suppressing the formation of conductive islands within the nanowire. Gaps in the gate-array, for enhancement-mode nanowires, then always act as barriers. As a possible explanation for this unconventional gate response we suggest surface trap states which efficiently screen the nanowire core from potential variations in the environment. 
We find comparable results in different device geometries where the gates access the nanowire from the side, indicating a cause seemingly independent on gate-design. The highly localized gate-action stands in contrast to other common III-V enhancement-mode nanowire systems such as thin InAs \cite{DorschNanotechnol2019,WangNanoLett2018} and InSb \cite{NadjPRL2012,Pribiag2013electrical}, but also Ge/Si core-shell nanowires \cite{HuNatNano2007} where comparable bottom-, top- and side-gate architectures enable the formation of DQDs.

\begin{figure*}
    \includegraphics{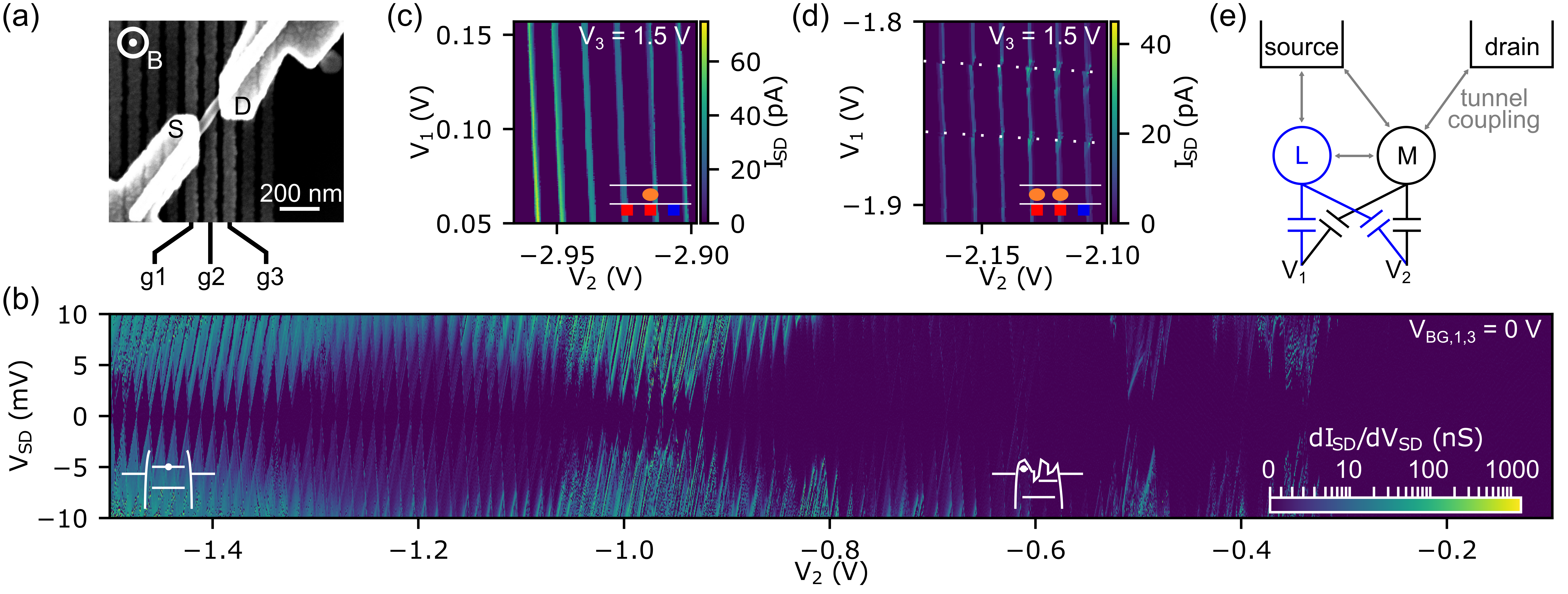}
    \caption{(a) Scanning electron microscope image of device B, a Schottky-contacted GaSb nanowire on three gate stripes g1 - g3. (b) Single-QD charge stability diagram at $V_{\mathrm{BG}}=0\,\si{V}$ and $V_{1,3}=0\,\si{V}$. Insets indicate the impact of the valence band edge. (c,d) Charge stability diagram at $V_{\mathrm{SD}}=1\,\si{mV}$, $V_{\mathrm{BG}}=0\,\si{V}$ as a function of $V_1$ and $V_2$ in a configuration where only g2 (c) or g1 and g2 (d) induce conductive islands in the nanowire. Insets indicate active gates (red) and QD formation. (e) Device configuration in (d).\label{fig2}}
\end{figure*}

Next, we study the effect of gates located directly under the Schottky contacts as well as the properties of QDs induced by local bottom-gating in the simplified single QD device B shown in Fig.~\ref{fig2}(a). We here reduce the number of bottom-gates (g1 - g3), where g1 and g3 are located below the Schottky contacts and g2 is designed to act as plunger gate and to induce a short QD in the enclosed nanowire segment.

The resulting charge stability diagram is shown in Fig.~\ref{fig2}(b). The data here is plotted as a function of the plunger gate voltage $V_2$ with all other gates grounded. For large negative voltages, $V_2 < -1.35\,\si{V}$, the charge stability diagram resembles that of a clean QD with clear diamond shaped Coulomb blockade regions. In contrast, for less negative $V_2$, additional overlaying patterns are observed and the Coulomb blockade regions are less well defined for lower occupancies. This is conceptually comparable to results obtained on p-type Si nanowires and are explained by a valence band roughness leading to multi-QD formation for low occupancies \cite{ZwanenburgJAppPhys2009}. This effect is indicated by the insets of Fig.~\ref{fig2}(b) and is expected to become less prominent and eventually vanish for decreased QD dimensions \cite{ZwanenburgJAppPhys2009}.

To study the effect of a voltage applied to g1 and g3, we next measure the charge stability diagram of the device as a function of $V_1$ and $V_2$ and find results following closely to the previously described gate functionality. For sufficiently positive $V_1$, the nanowire segment atop g1 remains non-conductive and the resulting charge stability diagram in Fig.~\ref{fig2}(c) shows Coulomb oscillations on the QD induced by g2, with a small cross-coupling to g1. In contrast, if $V_1$ is chosen more negative, a second conductive island is formed atop g1. In the resulting charge stability diagram, Fig.~\ref{fig2}(d), this manifests as steps in the Coulomb oscillations --- an indicator for Coulomb coupled QDs \cite{kellerPRL2016,ThierschmannNature2015,BohuslavskyiArXiv2020} where changes in occupancy of the left QD are marked by dashed white lines. Additionally, at the location of each step, finite bias triangles indicative of a DQD are observed \cite{VdWRMP2002}. We find the resulting system, sketched in Fig.\ref{fig2}(e) with all transport channels indicated, to be a hybrid between a serial and purely Coulomb coupled double quantum dot. Here, only the center QD is coupled to both contacts, while holes travelling through the left QD have to pass the middle QD to contribute to the detected current.

\begin{figure}
    \includegraphics{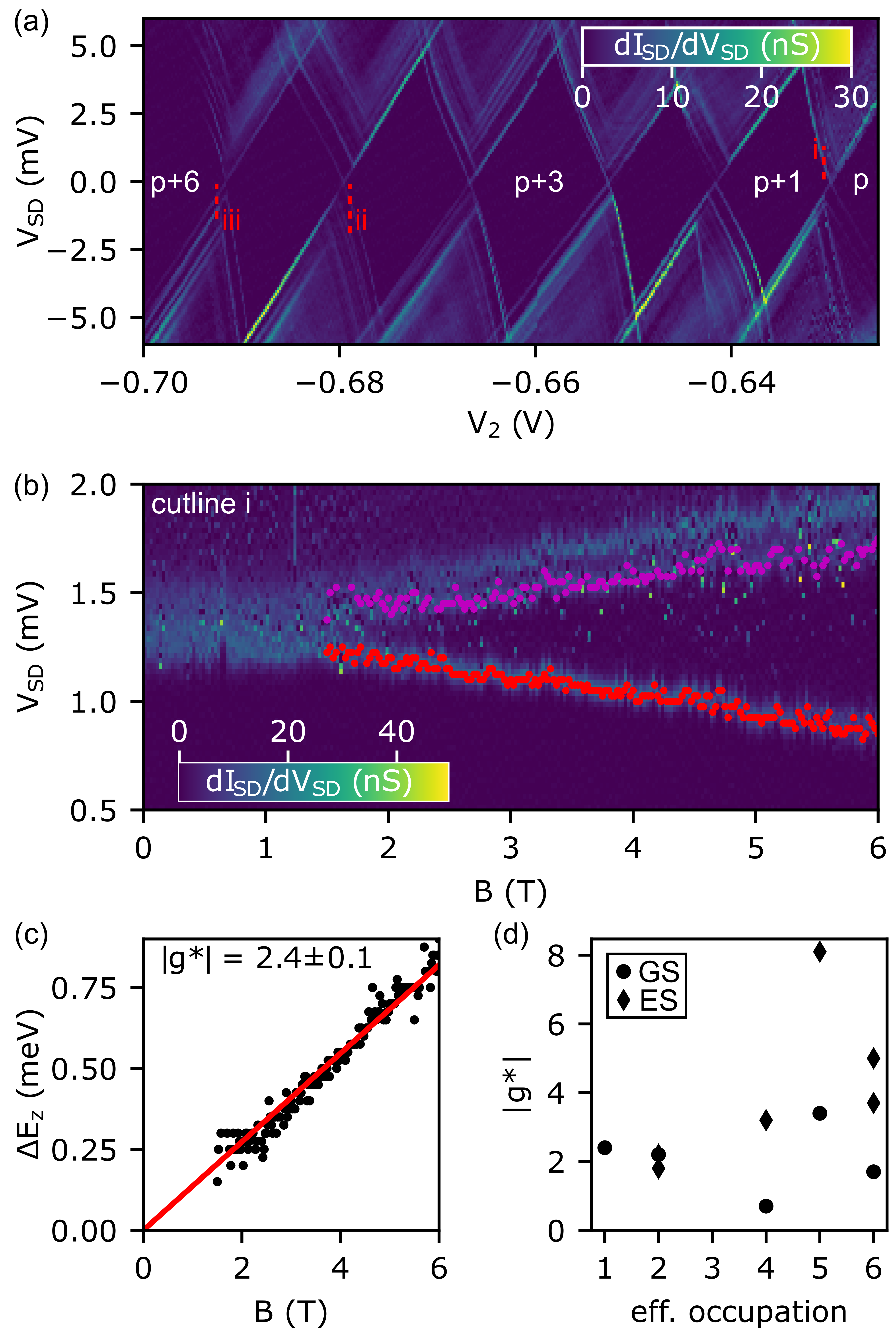}
    \caption{(a) Lowest six clean crossings in the charge stability diagram measured at $V_{\mathrm{BG}}=4\,\si{V}$ and $V_{1,3}=1.5\,\si{V}$. (b) Differential conductance along cutline i, labelled in (a) as a function of the magnetic field. Ground state Zeeman splitting is observed and conductance peaks corresponding to states with opposite spins are detected (red/magenta dots). (c) Fit to the Zeeman splitting $\Delta \mathrm{E}_\mathrm{z}$ extracted from (b). (d) $|g^*|$ for various ground (GS) and excited (ES) orbitals plotted against the effective occupation number of the QD, $\mathrm{p}=0$.\label{fig3}}
\end{figure}

We next apply a positive backgate $V_{\mathrm{BG}}= 4\,\si{V}$, $V_{1,3}=1.5\,\si{V}$ and a sufficiently negative $V_2$ to ensure the formation of exclusively one QD. The charge stability diagram for the lowest occupancies where the valence band edge roughness has no further visible impact on the electronic structure is shown in Fig.~\ref{fig3}(a) and we estimate a hole occupancy of $p\approx 25$.

To obtain information about effective $g^*$-factors of the QD, we perform magneto-transport spectroscopy on cutline i in Fig.~\ref{fig3}(a). The magnetic field is applied perpendicular to the nanowire, as illustrated in Fig.~\ref{fig2}(a). Fig.~\ref{fig3}(b) shows the differential conductance along cutline i as a function of the magnetic field, where Zeeman splitting of the ground state is observed. We note that the upper branch consists of a double conductance peak, which could be the result of nearly degenerate state  at $B=0\,\si{T}$ or a singlet and triplet state interacting at low magnetic fields below our resolution limit. The distance between data points on the upper and lower branch (purple and red markers) in Fig.~\ref{fig3}(b) follows a linear slope, see Fig.~\ref{fig3}(c), and in either of the cases represents closely the difference in energy between states of opposite spin. A fit (solid red line) with the standard expression for Zeeman splitting, $\Delta \mathrm{E}_\mathrm{z}=|g^*|\mu_\mathrm{B}B$ where $\mu_\mathrm{B}$ is the Bohr magneton, yields $|g^*|=2.4\pm 0.1$. This result corresponds well to the findings in GaSb/InAsSb core-shell nanowires in a hole transport regime \cite{GanjipourPRB2015}.

\begin{figure}
    \includegraphics{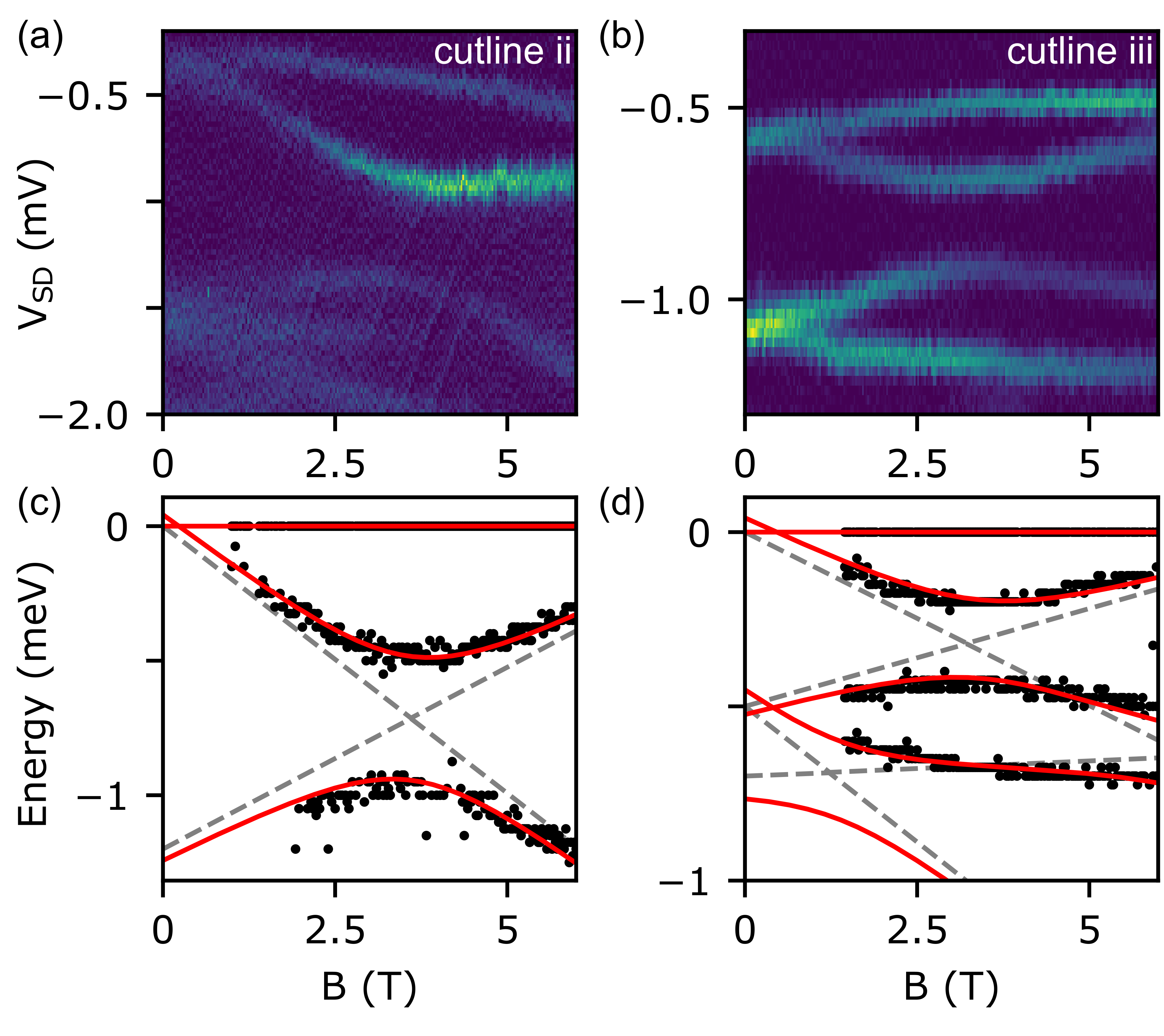}
    \caption{(a/b) Differential conductance along cutline ii/iii in Fig.~\ref{fig3}(a) as a function of the magnetic field. (c/d) Comparison and fit of datapoints extracted from (a/b) to calculated states with (solid red line) and without (grey dashed lines) level repulsion. All energies are normalized to the energetically highest ground state available for transport in the valence band.\label{fig4}}
\end{figure}

Magneto-transport spectroscopy measurements along cutlines near each of the crossings depicted in Fig.~\ref{fig3}(a) were analyzed and where Zeeman splitting is resolved $g^*$-factors are extracted. The resulting values found for ground- and excited state splitting are plotted against the effective occupation number in Fig.~\ref{fig3}(d). For an effective occupancy of $3$, the observed splittings are too small to clearly resolve and $|g^*|<0.2$ can be estimated. Overall, our data shows a spread in $g^*$-factors from almost vanishing up to $|g^*|=8.1\pm 0.2$ (see Fig.~\ref{fig4}(d)) with no clear trend between successive hole occupancies. 

For p-type nanowires, theory predicts large variations in $g^*$-factors between different orbital states \cite{CsontosPRB2007,CsontosAPL2008,CsontosPRB2008}, however that does not describe the absence of coinciding results between successive occupancies. We explain the observed variations within orbital states by the required plunger gate voltage difference between different occupancies. Variations of $V_2$ influence the dimensions of the conductive island within the nanowire and thus alter the wavefunctions leading to differences in the observed $g^*$-factors. 

An additional contribution to the $g^*$-factor variation is illustrated in Fig.~\ref{fig4}(a) and (b) where $dI_{\mathrm{SD}}/dV_{\mathrm{SD}}$ along the cutlines ii and iii is plotted as a function of the magnetic field. In both measurements Zeeman splitting of the ground and an excited state is observed and branches of opposite spin interact via the spin orbit interaction, leading to an avoided crossing. While in Fig.~\ref{fig4}(b) the two-fold spin degeneracy is obvious, only the state increasing in energy within the first excited orbital is clearly resolved in the data in Fig.~\ref{fig4}(a). We directly extract the spacings between the ground and first excited orbital $\Delta\epsilon_{0,1} = 1.2\,\si{meV}$ and $\Delta\epsilon_{0,1} = 0.5\,\si{meV}$ at $B=0\,\si{T}$ as well as the magnitude of the avoided crossing $2\Delta_{\mathrm{SO}} = 460\,\si{\mu eV}$ and $2\Delta_{\mathrm{SO}} = 225\,\si{\mu eV}$ from Fig.~\ref{fig4}(a) and (b) respectively. Here, based on the assumption that states remain energetically pinned by the strong coupling between the gate electrode and the QD upon variation of $V_{\mathrm{SD}}$, only a factor of two between the magnitude of the avoided crossing and the spin-orbit energy $\Delta_{\mathrm{SO}}$ is considered.

To obtain $g^*_n$-factor estimates for the ground ($n=0$) and first excited ($n=1$) orbital from Fig.~\ref{fig4}(a) and (b), we normalize the data to the energy of the ground state (state with the highest energy available for transport in the valence band) and compare the result to simple calculations in Fig.~\ref{fig4}(c) and (d). We thus describe the non-degenerate spin states by $\mathrm{E}_m = \pm\frac{1}{2}|g^*_n|\mu_\mathrm{B}B-\Delta\epsilon_{0,n}-\frac{1}{2}|g^*_0|\mu_\mathrm{B}B$ (grey dashed lines), where the last addend normalizes the energies with respect to the ground state and $\Delta\epsilon_{0,n}$ is the energetic spacing of orbital $n$ containing state $m$ to the highest available orbital. For states $m$,$m'$ in different orbitals and of different spin polarity, we further modify the energies by\begin{equation}
\mathrm{E}^{\mathrm{SO}}_{m,m'} = \dfrac{\mathrm{E}_m+\mathrm{E}_{m'}}{2}\pm\frac{1}{2}\sqrt{(\mathrm{E}_m-\mathrm{E}_{m'})^2+(2\Delta_{\mathrm{SO}})^2}\label{eq_levelRepulsion}
\end{equation}
to account for the spin-orbit interaction induced level repulsion \cite{GolovachPRB2008, FasthPRL2007}. 

By fitting Eq.~\ref{eq_levelRepulsion} (red lines) to the experimental data, we obtain $|g^*_0|=3.4\pm 0.2$ and $|g^*_1|=8.1 \pm 0.2$ as well as $|g^*_0|=1.7\pm 0.2$ and $|g^*_1|=3.7 \pm 0.2$ from Fig.~\ref{fig4}(c) and (d) respectively. While this is sufficient to reproduce the data in Fig.~\ref{fig4}(c), we note that the Zeeman split branches of the first excited states in (d) appear pinched. To reproduce this shape an additional state $\mathrm{E}_5$ is considered in Fig.~\ref{fig4}(d).

Finally, it is important to address the zero B-field Zeeman splitting for the calculations (solid red lines) presented in Fig.~\ref{fig4}(c) and (d), where a two-fold spin degeneracy of all states is expected. This zero-field splitting is a result of the approximative nature of Eq.~\ref{eq_levelRepulsion}, which only yields accurate results for $\Delta_{\mathrm{SO}}\ll \Delta \mathrm{E}_\mathrm{z}$ \cite{GolovachPRB2008}. Thus, the extracted $|g^*|$ values are to be considered reasonable estimates. Nevertheless, a comparison of the calculations including and excluding (solid red and dashed grey lines) spin-orbit interaction in Fig.~\ref{fig4}(c) and (d) illustrates clearly that a fit for $g^*$-factors as shown in Fig.~\ref{fig3}(c), even in a low B-field regime, underestimates the $g^*$-factor if avoided crossings occur. While this is taken into account in measurements where avoided crossings were obvious, those cases are not always clear within the experimental data. Consequently, the $|g^*|$ values in Fig.~\ref{fig3}(d) should be considered as lower bounds.

In conclusion, we demonstrated the formation of serial multi- and single-QDs in enhancement-mode Schottky-contacted GaSb:Zn nanowires. Our measurements show that bottom-gates can be used to form single and multiple QDs. However, we find that gates only act on the nanowire in their direct vicinity and can induce conductive islands while gaps in gate arrays result in barrier formation. This renders common gating approaches inept for the formation of controlled serial double quantum dots. As a solution, we propose device designs based on stacked gate arrays or a combination of bottom- and side- or top-gates. Further variations of the dopant concentration, nanowire diameter or surface treatment may also lead to improved gating behaviour.

Through magneto-transport spectroscopy and analyzing the observed avoided crossings, we estimate $g^*$-factors of up to $8.1\pm 0.2$ and a spin-orbit energy $\Delta_{\mathrm{SO}}$ reaching $230\si{\mu eV}$. Our findings for $\Delta_{\mathrm{SO}}$ are comparable to the results for electrons in III-V nanowire systems with effective $g^*$-factors reaching those commonly observed in InAs \cite{PfundPRB2007,FasthPRL2007,NadjPRL2012,NilssonNanoLett2009,WeperenPRB2015}. A direct comparison of the properties of GaSb nanowires with other p-type nanowire systems reveals $g^*$-factors matching or slightly exceeding those found for InSb \cite{Pribiag2013electrical}, Ge-hut \cite{ScappucciNatRevMat2020}, Si \cite{ZwanenburgNanoLett2009} and Si-Ge core-shell \cite{BraunsPRB2016} nanowires but $\Delta_{\mathrm{SO}}$ is up to an order of magnitude lower compared to selected Ge based systems \cite{ScappucciNatRevMat2020,HaoNanoLett2010,HigginbothamPRL2014}. The next challenge toward characterizing GaSb nanowire quantum devices will be to form a pure DQD and probe the various mechanisms affecting the hole spin relaxation time.

\section*{Acknowledgements}

The authors thank Timm M\"orstedt and Hanna Kindlund for contributions in the early stages of the project, Martin Leijnse and Adam J\"onsson for helpful discussions and acknowledge funding by the Swedish Research Council (VR) (project 2015-00619), the Marie Skłodowska Curie Actions, Cofund, Project INCA 600398, the Crafoord Foundation and by NanoLund. Device fabrication was carried out in the Lund Nano Lab (LNL).


%

\end{document}